\documentclass[eqsecnum,aps,twocolumn]{revtex4}
\usepackage{graphics}

\begin{document}
\title{\sc On Two Complementary Types of Directional Derivative and Flow Field Specification in Classical Field Theory}

\author{R.\ Smirnov-Rueda}
\address{Applied Mathematics Department, Faculty of Mathematics,
Complutense University, 28040, Madrid, Spain}

\date{\today}
\begin{abstract}
We discuss a general definition of directional derivative of any
tensor flow field and its practical applications in physics. It is
shown that both Lagrangian and Eulerian descriptions as
complementary types of flow field specifications adopted in modern
theoretical hydrodynamics, imply two complementary types of
directional derivatives as corresponding mathematical
constructions. One of them is the Euler substantive derivative
useful only in the context of initial Cauchy problem and the
other, called here as the local directional derivative, arises
only in the context of so-called final Cauchy problem. The choice
between Lagrangian and Eulerian specifications is demonstrated to
be equivalent to the choice between space-time with Euclidean and
Minkowski metric for any flow field domain, respectively.
Mathematical consideration is developed within the framework of
congruencies for general 4-dimensional differentiable manifold.
The analytical expression for local directional derivative is
formulated in form of a theorem. Although the consideration is
developed for one-component (scalar) flow field, it can be easily
generalized to any tensor field. Some implications of the local
directional derivative concept for the classical theory of fields
are also explored.
\end{abstract}
 \maketitle

\section{Introduction}

Physical background and motivation for Euler's mathematical
construction known as substantive (or material) derivative becomes
manifest in the way how the whole flow field may be specified in
hydrodynamics. Usually, two complementary types of specifications
(or representations) are thought to suffice in order to provide a
general description of the flow field kinematics. The first,
Lagrangian specification is based on identifying individual
elements or bits of fluid domain. This idea associates a fluid
motion with a geometrical transformation $H_{t}$ on the closure
$\bar{\Omega}_{0}$ such that the set $H_{t}\bar{\Omega}_{0}$
represents the same individual bit of fluid at time $t$. This
representation is valid only if the identification can be
maintained by some kind of labelling usually denoting the initial
position at instant $t_{0}$. The second, Eulerian specification
was conceived as dissociated from identification of individual
bits of fluid, only making use of the flow quantities as functions
of local position in space at each instant of time during the
motion. This specification results especially useful for
hydrodynamics of liquids and electromagnetic field description in
which any attempt of Lagrangian identification is impossible.
Thus, in Euler's approach the velocity vector field is a primary
notion. Assigning a velocity vector to each point of the fluid
domain, one obtains the system of ordinary differential equations
and their solutions as integral curves, intimately related to the
given velocity vector field. The Eulerian representation provides
a time parameterization of the curve in
local coordinate system as a differentiable mapping from an open set of $%
R^{1}$ into $R^{3}$.

Both complementary types of specifications (Eulerian and
Lagrangian), generally speaking, different and certainly useful in
complementary contexts, can be made mathematically equivalent
under special conditions. In terms of modern notation, they result
equivalent within the formulation of the initial Cauchy problem
for an ordinary differential equation for velocity field. It gives
a kind of dictionary for translating from one specification to the
other.

In this respect it is interesting to note that from the very birth
of theoretical hydrodynamics as an independent body of
mathematical knowledge, the conventional formulation of the
directional derivative of flow field quantities tacitly implies
the equivalence with the Lagrangian specification related to the
initial Cauchy problem. Thus, according to this type of
specification, Euler's substantial (or material) derivative
$\frac{Df}{Dt}$ describes the rate of time variation of
$f$-property of fluid element on its path from one to the other
point of space. As Euler himself coined it two and a half
centuries ago \cite{Kline}, his mathematical construction
described the rate of time variation of material properties
$following$ $the$ $motion$ $of$ $the$ $fluid$.

On the other hand, the conventional definition of the Lie
derivative on general differentiable manifolds, also admits the
same interpretation related to the initial Cauchy problem within
the framework of congruencies for a given parameterization
$\{\lambda _{i}\}$. When comparing scalars, vectors or tensors at
different points $\{\lambda _{i}\}$\ and $\{\lambda _{i}+\Delta
\lambda _{i}\}$\ on a certain integral curve, those entities under
comparison are Lie (or invariably) dragged back from $\{\lambda
_{i}+\Delta \lambda _{i}\}$ to the point $\{\lambda _{i}\}$. This
gives a unique difference and hence a unique derivative as the
limit of the difference between the values of scalars, vectors or
tensors at different points on a
manifold.\ A notion of a Lie invariably dragged function along a congruence $%
\lambda _{i}$\ is essential in the conventional definition of the
directional (or Lie's) derivative $\frac{d}{d\lambda _{i}}$ along
a flow (or general vector) field. It defines the rule necessary to
compare values of a given mathematical entity at two different
points.

The differential operator $\frac{d}{d\lambda _{i}}$ is a tangent vector to the curve $\mathbf{%
r}(\lambda _{i})$ on a manifold$.$ This association of the concept
of a directional derivative from the classical analysis allows to
maintain a $visual$ $picture$ on the Lie derivative as a tangent
vector that generates a kind of a $motion$ along the curve
$\mathbf{r}(\lambda _{i})$. In fact, when a differentiable
manifold is an ordinary Euclidean domain both Euler's and Lie's
derivatives coincide.

As we shall discuss in this paper, both Eulerian and Lagrangian
complementary flow field specifications which are traditionally
related to the initial Cauchy problem do not cover all possible
situations. It motivates a definition of complementary
mathematical object denoted here as a local directional
derivative. As a counter-part of the conventional approach, it
appears in this paper formulated within the
so-called final Cauchy problem and describes the time variation of $f$%
-property at some fixed point in a local coordinate system. It
allows the Euler description of the flow field to be dissociated
from any need of identification of individual bits of fluid and,
hence, to be considered properly as function of a position at
every instant of time.

\section{Euler's directional derivative and initial Cauchy problem}

    As the first task, let us explore the relationship between\ the Eulerian and
Lagrangian specifications within the conventional definition of
the directional derivative. To simplify our analysis, the further
discussion will be based on the consideration of one-component
(scalar) fluid moving in a three-dimensional Euclidean domain. We
denote by $f(\mathbf{r},t)$ some regular function in an arbitrary
space-time coordinate system. For instance, in fluid dynamics or
elasticity theory it could be a density of some physical medium
$\rho $ in Cartesian coordinates.

If in Lagrangian specification a geometrical transformation
$H_{t}$
represents a mapping of the closure $\bar{\Omega}_{0}$ onto $H_{t}\bar{\Omega%
}_{0}$ for the same individual bit of fluid at time $t$, then
$H_{t}$ also represents the function \cite{meyer}:

\begin{equation}
\label{odin} \mathbf{r}=H_{t}\mathbf{r}_{0}=\mathbf{r(r}_{0},t)
\end{equation}
where points $\mathbf{r}$ and $\mathbf{r}_{0}$ denote the
position-vector of the fluid identifiable point-particle at time
$t$ and initial time $t_{0}$, respectively. Thus, the velocity
vector field on the domain is defined as:

\begin{equation}
\label{dva}
\mathbf{v=}\frac{\partial }{\partial
t}\mathbf{r(r}_{0},t)
\end{equation}
where $\mathbf{r}_{0}$ is fixed.

In the context of Eulerian specification, there is no explicit
consideration of the function $\mathbf{r}=\mathbf{r(r}_{0},t)$.
The primary notion is the velocity field

\begin{equation}
\label{tri} \frac{d\mathbf{r}}{dt}=\mathbf{v(r},t)
\end{equation}
as function of position in space ($\mathbf{r)}$ and time ($t)$ on
a fluid domain. Based on this convention, it is clear that both
specifications
become mathematically equivalent when an initial condition $\mathbf{r}_{0}=%
\mathbf{r(}t_{0})$ for Cauchy problem is added to the ordinary
differential equation (\ref{tri}). It gives a rule for translating
from one specification
to the other. If any quantity has the Eulerian representation $f(\mathbf{r}%
,t)$, its Lagrangian representation is \cite{meyer}:

\begin{equation}
\label{chetyre} g(\mathbf{r}_{0},t)=f(\mathbf{r(r}_{0},t),t)
\end{equation}
and, therefore, Euler's material or substantive derivative is
conventionally defined as \cite{meyer}:

\begin{equation}
\label{pyat}
\frac{Df}{Dt}=(\frac{\partial }{\partial t}+(\mathbf{v},\mathbf{\nabla })%
\mathbf{)}f=\frac{\partial }{\partial t}g(\mathbf{r}_{0},t)
\end{equation}
where $\mathbf{r}_{0}$ is fixed.

The differential operator $\frac{Df}{Dt}$ has meaning only when
applied to flow field variables as functions of
$(\mathbf{r}(t),t)$ and gives the definition of directional
derivative as a time derivative $following$ $the$ $motion$ $of$
$the$ $fluid$ in the direction of its velocity field $\mathbf{v}$:

\begin{equation}
\label{shest}
\frac{D}{Dt}f(\mathbf{r}(t),t)=\lim\limits_{t\rightarrow
0}\frac{1}{t}\left[
f(\mathbf{r}(t_{0}+t),t_{0}+t)-f(\mathbf{r}(t_{0}),t_{0})\right]
\end{equation}
Here, according to the standard definition (\ref{pyat}) related to
the
initial Cauchy problem $\mathbf{r(}t_{0})=\mathbf{r}_{0}$, both values $%
f_{0}=f(\mathbf{r}(t_{0}),t_{0})$\ and
$f=f(\mathbf{r}(t_{0}+t),t_{0}+t)$ represent the $f$-property at
two different points of space $\mathbf{r}_{0}$ and
$\mathbf{r}_{0}+\Delta \mathbf{r}$, respectively.

Thus, in this type of Eulerian of specification, one can
reconstruct the
property $f$ of any identified bit of fluid at a new position $\mathbf{r}%
(t_{0}+dt)=\mathbf{r}_{0}+dr$ and instant $t_{0}+dt$, based only
on the knowledge of partial time derivative $\frac{\partial
f}{\partial t}$ and
local distributions of the gradient $\mathbf{\nabla }f$ and velocity field $%
\mathbf{v}$ in local coordinate system:

\begin{equation}
\label{siem} f=f_{0}+(\frac{\partial f}{\partial
t}+(\mathbf{v},\mathbf{\nabla }f)dt
\end{equation}

The function $f$ has, generally speaking, explicit as well as
implicit (through $\mathbf{r(}t)$) time dependencies and,
therefore, may be defined on a 4-dimensional space-time manifold
with no metric known $a$ $priori$. In this respect, the Lie
derivative, as a particularly interesting generalization of
(\ref{pyat}) (or (\ref{shest})), will be convenient for further
considerations, since it provides a necessary framework on
manifolds without metric. The
differential equation (\ref{tri}) will define a congruence or $t$%
-parameterized set of integral world-lines filling a 4-dimensional
manifold:

\begin{equation}
\label{vosem}
\frac{dx^{i}}{dt}=V^{i}\mathbf{;\qquad }x^{i}(t)=x_{0}^{i}\mathbf{+}%
\int\limits_{t_{0}}^{t_{0}+t}V^{i}dt
\end{equation}
where $x=(x^{0},x^{1},x^{2},x^{3})=(t,\mathbf{r})$; $V=(1,\mathbf{v})$; $%
x_{0}=x(t_{0})$ and, for our convenience, we leave for the time variable $%
x^{0}$ its original denomination $t$. Upper indices are used for
the coordinate functions $x^{i}(t)$ so that the 1-forms will
satisfy the index conventions of modern differential geometry.

If the velocity vector field $V$ is $C^{\infty }$, the coordinate
transformation (\ref{vosem}) is a diffeomorphism, forming part of
a one-parameter Lie group. Let us denote this transformation as $F_{t}$: $%
(x_{0}\rightarrow x(x_{0},t))$, which defines the mapping of $%
f(x_{0})=f(t_{0},\mathbf{r(}t_{0}))$ along the congruence (called
also as
Lie dragging \cite{schutz}) into a new function $f(x_{0}+x)=f(t_{0}+t,\mathbf{r}(t_{0}+t))$%
. A Lie dragging of scalar field has a simple geometrical
interpretation in Lagrangian specification of the fluid field: $F_{t}$ transforms the $f$%
-property of the identified fluid element at $x_{0}$ according the
rule \cite{dubrovin}:

\begin{equation}
\label{devyat} (F_{t}f)(x_{0})=f(F_{t}(x_{0}))=f(x_{0}+x)
\end{equation}
into the $f$-property of the same fluid element at $x_{0}+x$. This
interpretation also concerns the analytic expression of the Lie derivative $%
L_{V}$ along the vector field $V=(1,\mathbf{v)}$:

\begin{equation}
L_{V}\,f=[\frac{d}{dt}F_{t}\,f]_{t_{0}}=\lim\limits_{t\rightarrow 0}\frac{1}{%
t}[f(x_{0}+x)-f(x_{0})]  \label{desyat}
\end{equation}
where $x(t_{0})=0$. The concept of a Lie invariably dragged
function along a congruence is used in this conventional
definition (and implicitly in (\ref{shest})): in fact,
the quantity $f(x_{0}+x)$ is invariably dragged along the congruence from $%
x_{0}+x$ back to $x_{0}$, since in general tensor calculus it has
no clear meaning to compare both values $f(x_{0})$ and
$f(x_{0}+x)$ at different points in a manifold without metric.
Thus, (\ref{desyat}) gives a unique difference and therefore a
unique derivative. In fact, when $t$ is too small, the mapping
$F_{t}$: $(x_{0}\rightarrow x(x_{0},t))$ has an explicit form:

\begin{equation}
F_{t}x_{0}=x^{i}(x_{0},t)=x_{0}^{i}+tV^{i}(x_{0})+o(t)
\label{odinadzat}
\end{equation}
which gives analytic expression for the Lie derivative:

\begin{equation}
L_{V}\,f=\frac{d}{dt}f(F_{t}\,x_{0})=V^{i}\frac{\partial
f}{\partial x^{i}}
\label{dvenadzat}
\end{equation}

Important that both traditional definitions for Euler's and Lie's
directional derivatives, respectively, turn out to be defined
entirely in the spirit of original Lagrangian specification, i.e.
when a fluid element or a point on a congruence are constantly
identified in a local coordinate system. It explains why in an
ordinary Euclidean domain the Lie mathematical construction takes
a familiar form of Euler's directional derivative \cite{dubrovin}:

\begin{equation}
L_{V}\,f=V^{i}\frac{\partial f}{\partial x^{i}}=(\frac{\partial }{\partial t}%
+\mathbf{v\cdot \nabla )}f=\frac{Df}{Dt} \label{trinadzat}
\end{equation}
It is known as a full derivative along the vector filed
$V=(1,\mathbf{v)}$ and in hydrodynamics it also has numerous
applications for the description of the motion of macroscopic
individual bodies of fluid.

In this respect, let us now consider a time variation of the fluid $f$%
-content in a macroscopic 3-dimensional space domain $V$ moving
with the fluid:

\begin{equation}
\frac{d}{dt}\int\limits_{V(t)}fdV=\frac{d}{dt}\sum\limits_{\delta
V(t)}\int\limits_{\delta V(t)}fdV=\sum\limits_{\delta V(t)}\frac{d}{dt}%
(f\delta V)  \label{add1}
\end{equation}
where, for our convenience, the macroscopic volume $V(t)$ is
represented as the sum of individual microscopic volumes $\delta
V(t)$, i.e. $V=\sum \delta V$.

A geometrical transformation $H_{t}$ describes the evolution of
the whole volume $V(t)=H_{t}V_{0}$ as well as its individual bits
$\delta V(t)=H_{t}\delta V_{0}$. The assumption that the domain
$V$ and all $\delta V $\ move with the fluid, means that there is
no flux of $f$ through the common bounding surface of the fluid
domain $\partial V$ and bounding surfaces of all individual
elements $\partial \delta V$. It imposes the condition that normal
components of the relative fluid velocity field are zero at every
point of all such surfaces moving with the fluid. By the chain
rule we obtain from (\ref{add1}):

\begin{equation}
\frac{d}{dt}\int\limits_{V(t)}fdV=\int\limits_{V(t)}\frac{Df}{Dt}%
dV+\sum\limits_{\delta V(t)}\int f\frac{d}{dt}\delta V(t)
\label{add2}
\end{equation}

The time variation of a microscopic volume $\delta V$ is a result
of movement of each point of the bounding surface $\partial \delta
V$ and it is described by the divergence theorem \cite{batchelor}:

\begin{equation}
\frac{d}{dt}\delta V(t)=\int\limits_{\partial \delta V}(\mathbf{n},\mathbf{u}%
)dS=\int\limits_{\delta V}(\mathbf{\nabla },\mathbf{u})dV
\label{add3}
\end{equation}
where $\mathbf{n}$ is a unit vector normal to the surface
$\partial \delta V$ and $\mathbf{u}$ is the local velocity field
of surface points in the coordinate system attached to $\delta V$.
Since the divergence does not depend on the choice of coordinate basis, we can write $(\mathbf{\nabla },%
\mathbf{u})=(\mathbf{\nabla },\mathbf{v})$ for the flow velocity field $%
\mathbf{v}$ defined in the main coordinate system at rest. Hence
the local rate of expansion or dilation of a microscopic volume
element $\delta V$ is:

\begin{equation}
\frac{d}{dt}\delta V(t)=(\mathbf{\nabla },\mathbf{v})\delta V
\label{add4}
\end{equation}
When this is substituted into (\ref{add2}), one gets the result of the $%
Convection$ $Theorem$:

\begin{equation}
\frac{d}{dt}\int\limits_{V(t)}fdV=\int\limits_{V(t)}(\frac{Df}{Dt}+f(\mathbf{%
\nabla },\mathbf{v}))dV  \label{add5}
\end{equation}
More detailed and rigorous demonstrations can be found, for
instance, in \cite{meyer}, \cite{marsden}. We only need the way it
was reasoned as well as its interpretations for further
discussions.

To finish this Section, we conclude that the formula (2.13) for
Euler's derivative $\frac{Df}{Dt}$ $($as well as its integral
counter-part (\ref
{add5})) provide the rule for reconstruction of the $f$-property (or the $f$%
-content) at a new position $\mathbf{r=r}_{0}+d\mathbf{r}$ (or in
a new domain $V(t)=H_{t}V_{0})$, generally speaking, different
from the initial one. This hydrodynamics interpretation will
become useful in the next Section in order to contrast a
complementary specification of a flow field which provides the
reconstruction of the $f$-property at a fixed point of space
$\mathbf{r}_{0}$ but at instant $t_{0}+dt$ different from $t_{0}$,
entirely in the spirit of Euler's original idea of flow field
specification.

 \section{Final Cauchy problem and local directional derivative}

Let us discuss in this Section some limitations of conventional
definitions (\ref{shest}) and (\ref{desyat}) for a flow field in
situations when Lagrangian identification is in principle
impossible. Before proceed, it will be convenient to apply the
above-used terminology to clarify the definition of the partial
time derivative of a flow field $f$-quantity in the framework of
the classical analysis:

\begin{equation}
\frac{\partial f}{\partial t}=\lim\limits_{t\rightarrow
0}\frac{1}{t}\left[
f(\mathbf{r}_{0},t_{0}+t)-f(\mathbf{r}_{0},t_{0})\right]
\label{chetyrnadzat}
\end{equation}
where note that\ $\frac{\partial f}{\partial t}$ means not only
the rate of change of $f$ at a fixed position $\mathbf{r}_{0}$ in
space but (\ref
{chetyrnadzat}) is considered under the condition that the space variable $%
\mathbf{r}_{0}$ is not a subject of time parameter $t$ at all (i.e. $\mathbf{%
r}_{0}(t)=const$). In other words, this situation corresponds to
the time derivative of flow quantities with a $frozen$ velocity
field $\mathbf{v}$ (i.e. $\mathbf{v}=0$).

Let us now consider a time variation of the fluid $f$-content in a
fixed 3-dimensional space domain $V_{0}$. If we follow the way of
reasoning used earlier for the formulation of the $Convection$
$Theorem$, then:

\begin{equation}
\frac{d}{dt}\int\limits_{V_{0}}fdV=\frac{d}{dt}\sum\limits_{\delta
V_{0}}\int\limits_{\delta V_{0}}fdV=\sum\limits_{\delta V_{0}}\frac{d}{dt}%
(f\delta V_{0})  \label{abb1}
\end{equation}
where again, for our convenience, the macroscopic volume $V_{0}$
is represented as the sum of individual microscopic volumes
$\delta V_{0}$, i.e. $V_{0}=\sum \delta V_{0}$.

Since all volume elements $\delta V_{0}$ are now fixed, the
differential operator $\frac{d}{dt}$ acts only on $f$-property and
is used to be associated with the partial time derivative
$\frac{\partial }{\partial t}$, giving place to the well-known
relationship for a fixed domain $V_{0}$ (see any text on the
classical theory of fields):

\begin{equation}
\frac{d}{dt}\int\limits_{V_{0}}fdV=\int\limits_{V_{0}}\frac{\partial f}{%
\partial t}dV  \label{abb2}
\end{equation}

A note of caution is appropriate here. One might suspect that some
type of specification of the flow field for $f$-property inside
$\delta V_{0}$ is necessary as it was in (\ref{add2}) for the
reliable use of $\frac{Df}{Dt} $. The integrand in the left-hand
side of (\ref{abb1}) is not a simple multivariable function
$f(\mathbf{r},t)$ but a rather different mathematical entity
$f(\mathbf{r}(t),t)$, i.e. flow field quantity. This feature makes
it
difficult the straightforward application of the partial time derivative $%
\frac{\partial }{\partial t}$ meaningful for a fixed space variable $\mathbf{%
r}$ as function of a $frozen$ velocity field ($\mathbf{v}=0$).
Therefore, the use of the partial time derivative does not seem to
be fully justified in the conventional approach. This circumstance
was also critically pointed out in \cite{Chub}-\cite{Chubyk}.

To clarify the situation, let us go through the same mathematical
construction as was used before for Euler's and Lie's derivatives,
i.e we shall define an appropriate specification of the flow field
in this case. The terminology of ordinary differential equations
theory, combined with notions from classical analysis, enables us
to give a useful and compact definition of the left-hand side of (\ref{abb1}%
).

Let $\mathbf{r}_{0}$ be a fixed point of the closure $V_{0}$ and
let the path of some identified elementary bit of fluid (that at
some earlier instant $t_{0}$ passed through a certain point of
space $\mathbf{r}^{\ast }(t_{0})$) lie at present instant
$t_{0}+t$ on the position of the fixed point of space
$\mathbf{r}_{0}$. Then the full time derivative is understood as
the limit of the difference between the values of the volume
$f$-content at different instants $t_{0}$ and $t_{0}+t$. This
requirement provides a
framework necessary to derive analytic expression for the left-hand side of (%
\ref{abb1}):

\begin{eqnarray}
\label{shestnadzat}
\frac{d}{dt}\int\limits_{V_{0}}f(\mathbf{r}(t),t)dV=\nonumber \\
\lim\limits_{t%
\rightarrow 0}\frac{1}{t}\int\limits_{V_{0}}\left[ f(\mathbf{r}%
(t_{0}+t),t_{0}+t)-f(\mathbf{r}(t_{0}),t_{0})\right] dV
\end{eqnarray}
where $\mathbf{r}(t_{0})=\mathbf{r}_{0}$. Both values $f(\mathbf{r}%
(t_{0}),t_{0})$\ and $f(\mathbf{r}(t_{0}+t),t_{0}+t)$ represent the $f$%
-property at the same point of space $\mathbf{r}_{0}$.

Based on this convention we note that
$\mathbf{r(}t_{0}+t)=\mathbf{r}_{0}$ should lie at the end of the
integral curve:

\begin{equation}
\mathbf{r}_{0}=\mathbf{r}^{\ast }(t_{0})+\int\limits_{t_{0}}^{t_{0}+t}%
\mathbf{v}(\mathbf{r},t)dt
\label{siemnadzat}
\end{equation}

Extrapolation on a set of integral curves that fill our domain is
straightforward. The path of every such curve has to end at some
fixed point of the closure $V_{0}$. All of them are solutions of
initial Cauchy problems for the first-order differential equation:

\begin{equation}
\frac{d\mathbf{r}}{dt}=\mathbf{v};\qquad
\mathbf{r}(t_{0})=\mathbf{r}^{\ast }\in V^{\ast }
\label{vosemnadzat}
\end{equation}

In Lagrangian specification it represents a geometrical
transformation (or mapping) $H_{t}$ of the original 3-dimensional
domain $V^{\ast }$ at instant $t_{0}$ onto $H_{t}V^{\ast }=V_{0}$
at instant $t_{0}+t$. If the inverse mapping $H_{t}^{-1}$ is
single-valued, then $V^{\ast }=H_{t}^{-1}V_{0}$ can be regarded as
a reconstruction of initial conditions from the knowledge of the
final domain $V_{0}$, i.e. $V^{\ast }$ becomes dependent on $t$
parameter. Similarly to (\ref{odin}), for an individual bit of
fluid, $H_{t}$ defines the function:

\begin{equation}
\mathbf{r}_{0}=H_{t}\mathbf{r}^{\ast }=\mathbf{r(r}^{\ast
},t);\qquad \mathbf{r}_{0}\in V_{0}
\label{devyatnadzat}
\end{equation}
where $\mathbf{r}_{0}$ is fixed and belongs to $V_{0}$.

By analogy with the initial Cauchy problem for (\ref{tri}), it may
be called as a final Cauchy problem for (\ref{vosemnadzat}). If
any flow field quantity has the Eulerian representation
$f(\mathbf{r},t)$, its Lagrangian representation $g^{\ast }$ in
the context of the final Cauchy problem will be:

\begin{equation}
g^{\ast }(\mathbf{r}_{0},t)=f(\mathbf{r}(\mathbf{r}_{0},t),t)
\label{dvatzat}
\end{equation}
and, therefore, the partial time derivative $\frac{\partial g^{\ast }}{%
\partial t}$ (with the fixed $\mathbf{r}_{0}$)\ will define
some new mathematical construction:

\begin{equation}
\frac{\partial }{\partial t}g^{\ast }(\mathbf{r}_{0},t)=\frac{D^{\ast }f}{%
D^{\ast }t} \label{dvatzatodin}
\end{equation}

Note that the differential operator $\frac{D^{\ast }}{D^{\ast }t}$
makes sense only when applied to flow field variables as functions
of the final Cauchy problem and will be called here as the local
directional derivative by analogy with the definition of
$\frac{D}{Dt}$ within the framework of classical analysis:

\begin{equation}
\frac{D^{\ast }f}{D^{\ast }t}=\lim\limits_{t=0}\frac{1}{t}\left[ f(\mathbf{r(%
}t_{0}+t),t_{0}+t)-f(\mathbf{r}(t_{0}),t_{0})\right]
\label{dvatzatdva}
\end{equation}
where $\mathbf{r}(t_{0})=\mathbf{r}_{0}$ and we denote by
$\mathbf{r}(t_{0}+t)$ the final point $\mathbf{r}_{0}$ of the
integral curve (\ref {siemnadzat}).

To find the analytic expression for $\frac{D^{\ast }}{D^{\ast
}t}$, let us use a consideration similar to the applied in the
previous Section for the definition of $\frac{D}{Dt}$. The
differential equation (\ref{vosemnadzat}) will define a congruence
or $t$-parameterized set of integral world-lines filling a
4-dimensional space-time manifold:

\begin{equation}
\frac{dx^{i}}{dt}=V^{i}\mathbf{;\qquad }x_{0}^{i}(t)=x^{\ast i}\mathbf{+}%
\int\limits_{t_{0}}^{t_{0}+t}V^{i}dt
\label{dvatzattri}
\end{equation}
where $x^{\ast }=(t_{0},\mathbf{r}^{\ast })$ and $x_{0}=(t_{0}+t,\mathbf{r}%
_{0}).$

This transformation, which we denote as $G_{t}$: $(x^{\ast
}\rightarrow x_{0}=x(x^{\ast },t))$, defines the mapping of
$f(x^{\ast })$ along the congruence into a new function
$f(x_{0})$. In Lagrangian specification of the flow field it has
an obvious geometrical interpretation: $G_{t}$ transforms the $f$
property of the identified fluid element at $x^{\ast }$ according
the rule:

\begin{equation}
(G_{t}f)(x^{\ast })=f(G_{t}(x^{\ast }))=f(x_{0})
\label{dvatzatchetyre}
\end{equation}
The inconvenience of this description is that we are now at the
local coordinate system as a function of $\mathbf{r}^{\ast }(t)$.
Since the integration in (\ref{shestnadzat}) is effected over the
fixed domain $V_{0}$, we choose a local coordinate system attached
to $V_{0}$ by means of coordinate transformation:

\begin{equation}
\mathbf{r}^{\ast }(t)=\mathbf{r}_{0}-\int\limits_{t_{0}}^{t_{0}+t}\mathbf{v}(%
\mathbf{r},t)dt
\label{dvatzatpyat}
\end{equation}
or in 4-dimensional notations:

\begin{equation}
\frac{dx^{\ast i}}{dt}=V^{\ast i}\mathbf{;\qquad }x^{\ast i}(t)=x_{0}^{i}%
\mathbf{+}\int\limits_{t_{0}}^{t_{0}+t}V^{\ast i}dt
\label{dvatzatshest}
\end{equation}
where $V^{\ast }=(1,-\mathbf{v})$. It defines an equivalent
mapping which we denote as $G_{t}^{\ast }$: $(x_{0}\rightarrow
x^{\ast }=x(x_{0},t)).$

Note that $G_{t}^{\ast }$ is the same mapping $G_{t}$ but defined
in the local coordinate system attached to $V_{0}$. More
precisely, it means that the final Cauchy problem
(\ref{vosemnadzat}) admits an equivalent formulation as an
initial Cauchy problem (\ref{dvatzatpyat}) (i.e. $%
G_{t}^{\ast }$ is not the inverse transformation ($G_{t})^{-1}$,
since the course of time is not changed on the opposite one).

Thus, when $t$ is too small, the transformation $G_{t}^{\ast }$
has an explicit form:

\begin{equation}
G_{t}^{\ast }x_{0}=x^{\ast i}(x_{0},t)=x_{0}^{i}+tV^{\ast
i}(x_{0})+o(t)
\label{dvatzatsiem}
\end{equation}
which gives analytic expression for the Lie derivative along the congruence $%
V^{\ast }$ at a local coordinate system:

\begin{equation}
L_{V^{\ast }}\,f=\frac{d}{dt}f(G_{t}^{\ast }\,x_{0})=V^{\ast i}\frac{%
\partial f}{\partial x^{i}}
\label{dvatzatvosem}
\end{equation}
In an ordinary Euclidean domain this mathematical construction
takes the following expression:

\begin{equation}
L_{V^{\ast }}\,f=(\frac{\partial }{\partial t}-(\mathbf{v},\mathbf{\nabla })%
\mathbf{)}f=\frac{D^{\ast }f}{D^{\ast }t} \label{dvatzatdevyat}
\end{equation}

Now, noting that the integrand in the right-hand side of the
equation (\ref{shestnadzat}) is $\frac{D^{\ast }f}{D^{\ast }t}$
according to the definition (\ref{dvatzatdva}), we can proceed to
the formulation of our result (\ref {dvatzatdevyat}) as a theorem
proven in an ordinary Euclidean domain:

\newtheorem{LCT}{Theorem}
\begin{LCT}{(Local Convection Theorem):} Let $\mathbf{v}$ be a vector field
generating a fluid flow through a fixed 3-dimensional domain $V_{0}$ and if $%
f(\mathbf{r},t)\in C^{1}(\bar{V}_{0})$, then

\begin{equation}
\frac{d}{dt}\int\limits_{V_{0}}fdV=\int\limits_{V_{0}}(\frac{\partial }{%
\partial t}-(\mathbf{v},\mathbf{\nabla })\mathbf{)}fdV
\label{tridzat}
\end{equation}
where dV denotes the fixed volume element.
\end{LCT}
This result formulated within the framework of final Cauchy
problems could be regarded as a complementary counter-part of the
$Convection$ $Theorem$ (\ref{add5}) considered within the
framework of initial Cauchy problems.

In Eulerian specification, the formula (\ref{dvatzatdevyat})
admits a clear hydrodynamic interpretation: $\frac{D^{\ast
}f}{D^{\ast }t}$\ provides the
rule for reconstruction of the $f$-property at a fixed point of space $%
\mathbf{r}_{0}$ at instant $t_{0}+dt$, based only on the knowledge
of partial time derivative $\frac{\partial f}{\partial t}$ and
local distributions of gradient $\mathbf{\nabla }f$ and velocity
field $\mathbf{v}$ in the vicinity of $\mathbf{r}_{0}$:

\begin{equation}
f=f_{0}+(\frac{\partial f}{\partial t}-\mathbf{v\cdot \nabla }f)dt
\label{tridzatodin}
\end{equation}

Generally speaking, this type of Eulerian specification of the
flow field does not imply any sort of identification of fluid
elements and hence ought to be complementary to the original
Lagrangian approach. In fact, it considers the rate of time
variation of $f$-property locally, at fixed position of space.
Whereas the Euler derivative complementarily describes the rate of
time variation $following$ $the$ $motion$ $of$ $the$ $fluid$.

Both types of directional derivatives $\frac{Df}{Dt}$ and $\frac{D^{\ast }f}{%
D^{\ast }t}$ can be analyzed in terms of 1-forms or real-valued
functions of vectors in 4-dimensional manifolds:

\begin{equation}
\omega =(\omega _{i})=(\frac{\partial f}{\partial x^{i}})
\label{tridzatdva}
\end{equation}
where $i=0,1,2,3$ and $(\frac{\partial f}{\partial x^{i}})=(\frac{\partial f%
}{\partial t},\mathbf{\nabla }f)$ in an ordinary Euclidean domain.

Now we point out that in tensor algebra the set $\{\omega
_{i}V^{j}\}$ are components of a linear operator or $(
\begin{array}{c}
1 \\
1
\end{array}
)$ tensor. The formation of a scalar $\omega (V)$\ is called the
contraction of the 1-form $\omega $ with the vector $V$ and it is
an alternative representation of directional derivatives:

\begin{equation}
\frac{Df}{Dt}=\omega _{i}V^{i};\qquad \frac{D^{\ast }f}{D^{\ast
}t}=\omega _{i}V^{\ast i}  \label{tridzattri}
\end{equation}

The contraction of diagonal components of the tensor $\omega
_{i}V^{j}$\ is independent of the basis. Importantly, this law
shows that both types of directional derivatives $\frac{Df}{Dt}$
and $\frac{D^{\ast }f}{D^{\ast }t}$ are invariant and do not
depend on the choice of a local coordinate system. On the other
hand, it is also the property of scalar product in manifolds with
metric. The metric tensor maps 1-forms into vectors in a 1-1
manner. This pairing is usually written as:

\begin{equation}
\omega _{i}=g_{ij}\omega ^{j};\qquad V^{i}=g^{ij}V_{j}
\label{tridzat4}
\end{equation}
Therefore, from the point of view of tensor algebra,
(\ref{tridzattri}) can be considered as a scalar product in a
4-dimensional manifold with metric:

\begin{equation}
\frac{Df}{Dt}=g_{ii}\omega ^{i}V^{i};\qquad \frac{D^{\ast }f}{D^{\ast }t}%
=g_{ii}\omega ^{i}V^{\ast i}  \label{tridzat5}
\end{equation}
where $g_{ij}=\delta _{ij}$ is the Euclidean metric tensor. A
Minkowski
metric is also consistently singled out for local directional derivative $%
\frac{D^{\ast }f}{D^{\ast }t}$:

\begin{equation}
\frac{D^{\ast }f}{D^{\ast }t}=g_{ii}\omega ^{i}V^{\ast
i}=g_{ii}^{\ast }\omega ^{i}V^{i}  \label{tridzat6}
\end{equation}
where $V^{\ast }=(1,-\mathbf{v})$; $g_{ij}^{\ast
}=diag(1,-1,-1,-1)$ is indefinite or Minkowski metric tensor.

Another consequence of this form is that it gives orthonormal
bases for space-time manifolds (previously introduced with no
metric known $a$ $priori$). For Lagrangian flow field
specification, a basis is Cartesian and a transformation matrix
$\Lambda _{c}$ from one such basis to another is orthogonal
matrix:

\begin{equation}
\Lambda _{c}^{T}=\Lambda _{c}^{-1};\qquad ^{\prime }g_{ij}=\Lambda
_{c}^{-1}g_{ij}\Lambda _{c}  \label{tridzat7}
\end{equation}
These matrices $\Lambda _{c}$ form the symmetry group $O(4)$.

Likewise, for Eulerian specification a Minkowski metric picks out
a preferred set of bases known as pseudo-Euclidean or Lorentz
bases. A transformation matrix $\Lambda _{L}$ from one Lorentz
basis to another satisfies:

\begin{equation}
\Lambda _{L}^{T}=\Lambda _{L}^{-1};\qquad ^{\prime }g_{ij}^{\ast
}=\Lambda _{c}^{-1}g_{ij}^{\ast }\Lambda _{c}  \label{tridzat8}
\end{equation}
$\Lambda _{L}$ is called a Lorentz transformation and belongs to
the Lorentz group $L(4)$ or $O(3,1)$.

The point that needs to be emphasized here is the remarkable
circumstance of Euler's specification in evoking of the Minkowski
metric without any previous postulation. In other words,
consistent mathematical description of fluids is perfectly
compatible with the Lorentz symmetry group. This fact was not
seriously considered in theoretical hydrodynamics until now.

The Galilean group as one of subgroups of $O(4),$ is commonly used
in modern classical mechanics in flat space-time manifolds. This
is not surprising in view that all classical mechanics laws are
written in Lagrangian specification by constant identification of
mechanical objects and within the formulation of initial Cauchy
problem for equations of motion. It was therefore natural to admit
that space-time in classical mechanics has a Galilean group
symmetry. The Special Relativity postulation of Lorentz group
symmetry on mechanics is not trivial, having in mind the
complementary character of Lagrangian and Euler's descriptions.
Perhaps it can explain a paradoxical nature of some conclusions in
relativistic mechanics but it overcomes the scope of this work and
will be considered elsewhere.

Thus, from the point of view of flow field specifications, both
kinds of directional derivatives are complementary and equally
valid but should be used in different contexts. Euler's derivative
has therefore a more narrow framework of applicability in the
classical field theory than it was supposed. In what follows we
will confine our attention on some example from the classical
field theory.

\section{Local directional derivative in classical field theory}

Let us now consider the description of the conservation of the fluid $f$%
-content in an arbitrary 3-dimensional space domain. If the volume
$V$ moves with the fluid, the $Convection$ $Theorem$ (2.18)
written in differential form states that the $f$-content is
conserved when the total time derivative is zero:

\begin{equation}
\frac{Df}{Dt}+f(\mathbf{\nabla },\mathbf{v})=\frac{\partial f}{\partial t}+%
\mathbf{\nabla }(f\mathbf{v})=0  \label{acc1}
\end{equation}
In particular, when the velocity field $\mathbf{v}$ is locally
zero, it represents the continuity equation of any elastic medium
locally at rest:

\begin{equation}
\frac{\partial f}{\partial t}+f(\mathbf{\nabla },\mathbf{v})=0
\label{acc2}
\end{equation}
where the extra term ($\mathbf{v,\nabla }f)$ due to the fluid
movement has disappeared.

In the case when the fluid moves through a volume $V_{0}$ fixed in
local coordinate system at rest, a mathematical restriction on
conservation immediately leads to the well-established
integro-differential form of continuity equation:

\begin{equation}
\frac{d}{dt}\int\limits_{V_{0}}fdV=-\oint\limits_{S_{0}}f(\mathbf{v\cdot dS}%
)=-\int\limits_{V_{0}}\mathbf{\nabla }(f\mathbf{v})dV
\label{tridzat9}
\end{equation}
Note that here both sides of the equation are obviously
independent on the choice of a particular coordinate basis.

As we already mentioned earlier in the previous Section, it is
commonly thought that, in this case, the total time derivative can
be substituted in the integrand by the partial derivative, giving
place to the conventional form of continuity equation in the
reference system at rest:

\begin{equation}
\frac{\partial f}{\partial t}=-\mathbf{\nabla }(f\mathbf{v})
\label{sorok}
\end{equation}
The circumstance that it coincides with the expression
(\ref{acc1}) derived for the volume in motion, is mainly
attributed to the cross-verification of the standard differential
form of continuity equation. Nevertheless, it is strange to
contemplate that the differential equation (\ref{sorok}) does
not possess the symmetry properties of its original integral counter-part (%
\ref{tridzat9}). The left-hand side of (\ref{sorok}) becomes
manifestly dependent on the choice of a coordinate basis which,
generally speaking, leads to a more narrow group of symmetries.
Let us see whether results of the previous Section may help to
clarify the situation.

In fact, implementation of Euler's type of flow field
specification for the left-hand side of (\ref{tridzat9}) in the
framework of the final Cauchy problem changes the character of the
integrand expression. If it is considered as the local directional
derivative of $f$-property, the continuity equation (%
\ref{tridzat9}) takes the following differential form:

\begin{equation}
\frac{D^{\ast }f}{D^{\ast }t}=\frac{\partial f}{\partial t}-(\mathbf{v},%
\mathbf{\nabla })f=-\mathbf{\nabla }(f\mathbf{v})  \label{sorok1}
\end{equation}
that coincides with (\ref{acc2}). The right-hand side of
(\ref{sorok1}) as a
divergence and the left-hand side as the local directional derivative $\frac{%
D^{\ast }}{D^{\ast }t}$ do not dependent on the choice of a
coordinate basis. It means that this differential form of the
continuity equation has
the symmetry properties of its original integral counter-part (\ref{tridzat9}%
).

A brief comment is worthy in this respect. Why the traditional
approach based on the $Convection$ $Theorem$ gives a different
result (\ref{acc1})?
Certainly, it is correct but it has a non-invariant extra term ($\mathbf{%
v,\nabla }f)$ due to the fact that the description is effected in
the
reference system at rest for the domain $following$ $the$ $motion$ $of$ $the$ $%
fluid$. These shortcomings of the direct application of the $Convection$ $%
Theorem$ was not appreciated until now. If an observer moves with
the fluid, this $Theorem$ gives the equation (\ref{acc2}) without
an extra term and with the symmetry properties of the original
integral equation (\ref {tridzat9}). On the other hand, one could
logically ask why all numerical simulations based on the standard
differential form of the continuity equation (\ref{sorok})\ do not
lead to incorrect predictions? The answer is the following:
traditional time discretization schemes for the partial time
derivative of flow field quantities (see, for instance,
\cite{iserles}) treat it as if it were the total time derivative.

Another interesting task would be an application of the local
derivative concept to the integral form of Maxwell's equations.
Two of them contain the full time derivative over volume integrals
and are known as induction laws for electric $\mathbf{E}$ and
magnetic $\mathbf{H}$ vector fields, respectively, in the local
frame of reference at rest:

\begin{equation}
\int\limits_{C}(\mathbf{H},d\mathbf{l})=\frac{4\pi }{c}\int\limits_{S}(%
\mathbf{j},d\mathbf{S})+\frac{1}{c}\frac{d}{dt}\int\limits_{V}(\mathbf{%
\nabla },\mathbf{E})dV  \label{sorok2}
\end{equation}

\begin{equation}
\int\limits_{C}(\mathbf{E},d\mathbf{l})=-\frac{1}{c}\frac{d}{dt}%
\int\limits_{V}(\mathbf{\nabla },\mathbf{H})dV  \label{sorok3}
\end{equation}

Straightforward application of (\ref{tridzat}) in this case is
hindered by $a$ $priori$ unknown nature of the velocity vector
field for electric and magnetic field components. At this stage,
only quasistatic approximation can admit a reliable application of
the local directional derivative concept. In fact, the Special
Relativity firmly established that electromagnetic field
components of uniformly moving single charge do not depend
explicitly on time parameter $t$. In other words, $\mathbf{E}$ and
$\mathbf{H}$ are thought to be rigidly attached to the charged
particle and uniformly move with it. This is one of the
consequences of the Relativity Principle. Thus, if the charge
velocity $\mathbf{v}_{q}$ is known, the velocity vector field
$\mathbf{v}$ for quasistatic components of electric and magnetic
field is also defined in the closure $V$. Applying the result of
the Theorem (\ref{tridzat}), we can rewrite
(\ref{sorok2})-(\ref{sorok3}) in a more convenient form:

\begin{equation}
\int\limits_{C}(\mathbf{H},d\mathbf{l})=\frac{4\pi }{c}\int\limits_{S}(%
\mathbf{j},d\mathbf{S})+\frac{1}{c}\int\limits_{V}(\frac{\partial
}{\partial t}-(\mathbf{v},\mathbf{\nabla }))\mathbf{\nabla E}dV
\label{sorok4}
\end{equation}

\begin{equation}
\int\limits_{C}(\mathbf{E},d\mathbf{l})=-\frac{1}{c}\int\limits_{V}(\frac{%
\partial }{\partial t}-(\mathbf{v},\mathbf{\nabla }))\mathbf{\nabla H}dV
\label{sorok5}
\end{equation}
where $\mathbf{v=v}_{q}$ is the instantaneous velocity field in the closure $%
V$.

Since the motion is uniform, all partial derivatives vanish from
(\ref {sorok4})-(\ref{sorok5}). Applying a well-known expression
for a general vector field $\mathbf{A}$:

\begin{equation}
(\mathbf{v},\mathbf{\nabla })\mathbf{A}=\mathbf{v}(\mathbf{\nabla A})-[%
\mathbf{\nabla },[\mathbf{v},\mathbf{A}]]  \label{sorok6}
\end{equation}
and reducing the volume $V$ to zero, we arrive to the
well-established relationship between quasistatic magnetic and
electric field strength of an uniformly moving charge from the
point of view of a local inertial reference system \cite{landau}:

\begin{equation}
\mathbf{H}=\frac{1}{c}[\mathbf{v},\mathbf{E}];\qquad \mathbf{E}=-\frac{1}{c}[%
\mathbf{v},\mathbf{H}]  \label{sorok7}
\end{equation}

It is worth stressing that $a$ $priori$ no relativity principle
was needed in deriving these transformation rules for electric and
magnetic field components. The term proportional to
$-(\mathbf{v},\mathbf{\nabla })$ can be considered as convective
displacement current \cite{smir} by analogy with
Maxwell's displacement current proportional to $\frac{\partial }{\partial t}$%
. Note that the integral form of Maxwell's equations
(\ref{sorok4})-(\ref {sorok5}) written in Euler's specification is
now compatible with the charge conservation law (\ref{sorok1})
also represented in Euler's specification.

On the other hand, the Lorentz and Ampere force conceptions are
manifestly valid quasistatic approximations (\ref{sorok7}) and
therefore are inclosed into integral form of Maxwell's field
equations on a basic level. It means that there may be no need to
postulate them separately as it was done in Maxwell-Lorentz
microscopic electron theory and remains accepted at present.
Nevertheless, any full analysis of these issues comes out of the
scope of the present consideration and will be given elsewhere.

\section{Conclusions}

We attempted to consider a logical background and structure of
useful mathematical constructions which are traditionally based on
both Eulerian and Lagrangian flow field representations,
complementary to each other. This account provides a mathematical
method that justifies the definition of a complementary
counter-part for Euler's directional derivative which is called
here as the local directional derivative.

The point that needs to be emphasized is the complementary
character of the above introduced concept. By no means it
substitutes the Euler mathematical construction. By contrary, it
is shown that both types of directional derivatives are equally
valid but should be used in different contexts. In fact, Euler's
substantive derivative arises in the context of initial Cauchy
problems and therefore becomes useful within the framework of the
Lagrangian type of description of flow field quantities. Likewise,
it is possible to define a complementary framework of so-called
final Cauchy problems appropriate for the Euler flow field
specification as a function of position in space and in time for
fluid domain. From the point of view of the classical theory of
fields it means a more narrow framework of applicability for
Euler's derivative than it was thought.

The analytic expression for the local directional derivative is
formulated
in form of a theorem analogous and complementary to the $Convection$ $%
Theorem $ well-established in theoretical hydrodynamics. One of
its interesting conclusions is that the choice between Lagrangian
and Eulerian types of flow field representation is equivalent to
the choice between space-time manifolds with Euclidean and
Minkowski metric, respectively. Therefore, the consistent
mathematical description of kinematics of fluids in Eulerian
representation results compatible with the Lorentz group symmetry
$L(4)$. In fact, it could be understood as complementary to the
$O(4)$ group symmetry compatible with the Lagrangian
representation.

On the other hand, the definition of the mathematical construction
complementary to the traditional one, helps to get a deeper
insight on the cross-verification of several partial differential
equations obtained from their well-established integral
counter-parts in classical theory of fields.

Although the consideration in this work was developed for
one-component (scalar) flow field, the notion of the local
directional derivative can be easily generalized on Lie's
derivatives for any general tensor field on differentiable
manifolds. Both types of Lie's derivative will correspond to both
complementary types of specifications.

In place of concluding remark let us give asserting and
encouraging words of a great mathematician. Gauss once wrote in
his letter to Bessel (quoted form \cite{morris}): ...$One$
$should$ $never$ $forget$ $that$ $the$ $function$ [$of$ $complex$
$variable$], $like$ $all$ $mathematical$ $constructions$, $are$
$only$ $our$ $own$ $creations$, $and$ $that$ $when$ $the$
$definition$ $with$ $which$ $one$ $begins$ $ceases$ $to$ $make$
$sense$, $one$ $should$ $not$ $ask$, $what$ $is$, $but$ $what$
$is$ $convenient$ $to$ $assume$ $in$ $order$ $that$ $it$ $remain$
$significant$...".

\begin{acknowledgments}
 The author is indebted to his friends and colleagues who stimulated ideas
which make up this work, in particular, to Prof. A.E. Chubykalo,
Prof. V. Kassandrov and Dr. V. Onoochin.
\end{acknowledgments}


\end{document}